\documentclass[12pt]{spieman}  % 12pt font required by SPIE;
\usepackage{amsmath,amsfonts,amssymb}
\usepackage{graphicx}
\usepackage{setspace}
\usepackage{tocloft}
\usepackage{eurosym}
\usepackage{lineno}
\usepackage{booktabs}
\usepackage{tabularx}
\usepackage{xcolor}
\usepackage{array}
\usepackage{hyperref}
\newcolumntype{L}[1]{>{\raggedright\arraybackslash}p{#1}}

\title{A Paired Point-of-Care Ultrasound Dataset for Image Quality Enhancement and Benchmarking via a cGAN Baseline}
\author[a,b*,\textdagger]{Lennard M. van Karnenbeek}
\author[b\textdagger]{Hilde G.A. van der Pol}
\author[b]{Mark Wijkhuizen}
\author[c,e]{Eva Poelman}
\author[b,e]{Caroline A. Drukker}
\author[a,b,e]{Theo Ruers}
\author[b]{Freija Geldof}
\author[b,e]{Behdad Dashtbozorg}

\affil[a]{Department of Nanobiophysics, Faculty of Science and Technology, University of Twente, Drienerlolaan 5, 7522 NB Enschede, The Netherlands}
\affil[b]{Image-Guided Surgery, Department of Surgery, Netherlands Cancer Institute, Plesmanlaan 121, 1066 CX Amsterdam, The Netherlands}
\affil[c] {Department of Radiology, Netherlands Cancer Institute, Amsterdam, The Netherlands}
\affil[d] {Department of Medical Imaging, Radboud University Medical Center, Nijmegen, The Netherlands}
\affil[e] {Center for Early Cancer Detection, Netherlands Cancer Institute, Plesmanlaan 121, 1066 CX Amsterdam, The Netherlands}

\cftpagenumbersoff{figure}
\cftpagenumbersoff{table} 
\begin{document} 
\maketitle

% MAX 250 words. Can still add bit 
%Might need to re shape it to purpose,Approach, results, conclusion
\begin{abstract}\\
\textbf{Purpose:} We aim to enhance the image quality of point-of-care ultrasound (POCUS) devices using deep learning and a novel paired dataset of POCUS and high-end ultrasound
images. 
\\
\textbf{Approach:}
We collected the first accurately paired dataset using a custom-built automated gantry system of low-end POCUS and high-end ultrasound images. A conditional generative adversarial network (cGAN) was utilized based on the pix2pix architecture, with a U-Net generator that incorporates both L1 and structural similarity index (SSIM) losses to improve perceptual quality. Pretraining on a simulation dataset further boosts performance. Evaluation was performed on 1064 paired ex vivo tissue and phantom ultrasound image sets.
\\
\textbf{Results:}
Our approach improves the SSIM from
0.29 to 0.54 and PSNR from 19.16 dB to 22.41 dB. No-reference metrics also indicate substantial enhancement, with 
the Natural Image Quality Evaluator (NIQE) and Perception-based Image Quality Evaluator (PIQE) scores dropping from 7.95 to 4.44 and 31.12 to 19.99, respectively.
\\
\textbf{Conclusions:}
This work presents the first publicly available accurately paired dataset of low-end POCUS to high-end ultrasound images. Additionally, our results demonstrate the potential of the proposed framework to
overcome hardware limitations of handheld POCUS, enhancing its diagnostic value in low-resource and point-of-care settings.
The POCUS-IQ Dataset is publicly available at \url{https://github.com/NKI-MedTech-AI/POCUS-IQ}.
\end{abstract}

%\begin{abstract}
%Handheld point-of-care ultrasound (POCUS) devices are increasingly used due to their portability and affordability, yet they suffer from lower image quality compared to high-end systems. In this paper, we propose a deep learning-based framework to enhance the quality of POCUS images using a conditional generative adversarial network (cGAN). A key contribution of this work is the creation of the first accurately paired dataset of low-cost POCUS and high-end ultrasound images, acquired using a custom-built automated gantry system. A network was utilized based on the pix2pix architecture, with a U-Net generator that incorporates both L1 and structural similarity index (SSIM) losses to improve perceptual quality. Pretraining on a simulation dataset further boosts performance. Evaluation was performed on 1064 paired ex vivo tissue and phantom ultrasound image sets. Our approach improves the SSIM from 0.29 to 0.54 and PSNR from 19.16 dB to 22.41 dB. No-reference metrics also indicate substantial enhancement, with the Natural Image Quality Evaluator (NIQE) and Perception-based Image Quality Evaluator (PIQE) scores dropping from 7.95 to 4.44 and 31.12 to 19.99, respectively. 
% Qualitative assessment by ultrasound experts confirmed preference for enhanced images over the original POCUS images. 
%Our results demonstrate the potential of the proposed framework to overcome hardware limitations of handheld POCUS, enhancing its diagnostic value in low-resource and point-of-care settings.
%\end{abstract}
% Include a list of up to six keywords after the abstract

% Include email contact information for corresponding author
{\noindent \footnotesize\textbf{*}Address all correspondence to Lennard van Karnenbeek,  \linkable{l.v.karnenbeek@nki.nl}} \\
{\noindent \footnotesize\textbf{\textdagger}These authors contributed equally to this work} %Mark Wijkhuizen, \linkable{m.wijkhuizen@nki.nl}; Theo Ruers, \linkable{t.ruers@nki.nl}; Freija Geldof, \linkable{f.geldof@nki.nl}; Behdad Dashtbozorg, \linkable{b.dasht.bozorg@nki.nl}}

\begin{spacing}{2}   % use double spacing for rest of manuscript

\section{Introduction}
\label{sect:intro}  % \label{} allows reference to this section
The use of handheld point-of-care ultrasound (POCUS) devices has been on the rise in recent years. This increase in popularity can be attributed to several key characteristics of these devices. Firstly, their portability offers greater convenience compared to conventional cart-based devices. Moreover, these handheld POCUS devices are more affordable than traditional high-end ultrasound machines \cite{Hashim_21,gilbertson_20,Han_19,Pol_24}, making ultrasound technology more accessible and expanding its application beyond the radiology department in hospitals. For example using ultrasound intraoperatively\cite{Veluponnar_23,Geldof_23} or in a daily clinical setting\cite{NATALI_2025112274,Kurucz_2025}. POCUS is particularly valuable in settings where larger, more expensive ultrasound equipment is impractical, such as in bedside emergency care, general practitioner offices, home care environments, and rural medicine facilities, potentially reducing the pressure on hospital care.~\cite{Zhou_19,stewart_20,osterwalder_23, becker_16, evangelista_16}

Despite these advantages, POCUS typically suffers from reduced image quality compared to high-end systems. This is primarily caused by hardware limitations, including lower-grade transducers, limited processing power, and the absence of advanced post-processing algorithms \cite{henderson_17,ahn_15,Zhou_19,salimi_22}. As a result, POCUS images often exhibit reduced resolution and contrast, increased noise levels, and less distinct anatomical structures \cite{Khan_21}. These factors can negatively impact diagnostic accuracy and reduce clinician confidence, particularly in challenging imaging scenarios or for less experienced users\cite{Riley_14,Stock_15}. Despite the advancements in POCUS technology in recent years, a trade-off remains between image quality and the benefits of low cost and portability \cite{Zhou_21, jafari_20, henderson_17}.

%Improving image quality can be approached via hardware upgrades, enhanced beamforming, or post-processing techniques. However, hardware solutions increase cost and reduce portability, and raw radio frequency data needed for beamforming is rarely accessible. This motivates the use of deep learning for post-processing-based enhancement. Recent literature demonstrates the promise of these approaches, yet many rely on paired datasets with spatial misalignments that impair training and evaluation \cite{Zhou_19,Zhou_21,miccai_23}. A recent review of the literature highlights this issue and shows that simulation-based datasets, while useful for pretraining, may not adequately capture the variability encountered in clinical practice \cite{Pol_24}.

Improving image quality can be approached via hardware upgrades, enhanced beamforming strategies, or post-processing techniques. \cite{lockwood_98, Matrone_14, anaya_21, zhang_20,liu_23} However, hardware-based solutions increase system cost and reduce portability, while access to raw radio-frequency data needed for beamforming is rarely available on commercial POCUS devices. This motivates the growing interest in post-processing-based image enhancement methods.
Traditional post-processing approaches, including filtering and deconvolution, have demonstrated some success in improving the quality of ultrasound images. \cite{CONTRERASORTIZ2012419} However, recently, deep learning-based techniques have emerged as a promising alternative, achieving state-of-the-art performance in image quality enhancement tasks. 

Although recent studies highlight the potential of deep learning for improving POCUS image quality, many existing approaches rely on paired datasets that suffer from spatial misalignments between low-quality and high-quality images, which impairs training and evaluation. \cite{Zhou_19,Zhou_21,miccai_23} These misalignments arise due to the challenges in acquiring the exact same anatomical information using different probes and imaging systems. Registration techniques can be used to partially mitigate these misalignments, but registration alone is unable to fully resolve this issue. Moreover, misalignments introduce uncertainties in the ground truth, making it difficult to properly validate the quality enhancements achieved by these deep learning models. Although simulation data could offer an alternative solution, a recent literature review highlights that simulation-based datasets, while useful for pretraining, may not adequately resolve the issue. \cite{Pol_24} The higher performance gains observed in simulation-based models compared to in-vivo and phantom datasets suggest that such data may not fully capture the variability encountered in clinical practice.

%To address this, we collect a novel, accurately aligned paired dataset of low-cost POCUS and high-end ultrasound images, acquired using a custom motorized setup. We further develop a conditional GAN (cGAN) based on the pix2pix framework, integrating L1 and SSIM losses. This approach significantly improves image quality while preserving the practical advantages of POCUS, offering a promising direction for enhancing diagnostic confidence in low-resource and point-of-care settings.

To address these limitations in the literature, we collect a novel, accurately aligned paired dataset of low-cost POCUS and high-end ultrasound images of surgical specimens and an abdominal phantom. Alignment is ensured by using a custom motorized acquisition setup that precisely translates both probes over the same spatial location, in combination with a dedicated phantom to eliminate any prior spatial misalignments of the probes. Using this well-aligned paired dataset (called POCUS-IQ), we develop a supervised deep learning approach to translate low-quality POCUS images into high-quality ultrasound images. Specifically, we develop a conditional generative adversarial network (cGAN) based on the pix2pix framework, integrating L1 and SSIM losses.\cite{goodfellow_14,islam_24, nayak_24, jeong_22,Isola_16} This approach improves POCUS image quality without compromising its low-cost and portable nature, offering a promising direction for enhancing diagnostic confidence in low-resource and point-of-care settings. Moreover, this paper can be regarded as a stepping stone towards more sophisticated deep-learning models using this accurately paired dataset.

\section{Materials and Methods}

\subsection {POCUS-IQ dataset}
This section describes the acquisition of the paired ultrasound dataset using a low-cost POCUS device and a high-end ultrasound system. It outlines the acquisition setup and protocol, the specimens included, and the preprocessing steps applied prior to model training.

\label{sec2.1}
\subsubsection{Ultrasound devices}
To develop a paired dataset for image quality enhancement, images were acquired using two ultrasound systems: a low-cost handheld device (Telemed MicrUS EXT-1H, Vilnius, Lithuania) combined with an L15-6L25S-3 transducer, and a high-end cart-based system (Philips CX50, Eindhoven, The Netherlands) with an L15-7io transducer. An overview of the ultrasound devices is shown in Table \ref{device_specifications}. Compared to the Philips system, the Telemed device is significantly smaller, lighter, and more affordable, but has lower spatial resolution and fewer transducer elements. Both devices acquired images using a frequency of 14 MHz. In contrast to the Philips system, the Telemed system provided access to unprocessed ultrasound images. 

\begin{table}[htbp]
    \centering
    \small
    \setlength{\tabcolsep}{6pt}
    \renewcommand{\arraystretch}{1.2}
    \caption{Specifications of ultrasound devices.}
    \label{device_specifications}
    \begin{tabular}{@{}L{3cm} c c @{}}
        \toprule
        & \textbf{Telemed} & \textbf{Philips} \\
        \midrule
        Description
        & Handheld, wired, cost-effective
        & Cart-based, high-end, wired \\

        Device
        & MicrUS EXT-1H
        & CX50 \\

        Manufacturer
        & TELEMED, Lithuania
        & Philips, Netherlands \\

        Probe
        & TL15-6L25S-3
        & L15-7io \\

        Dimensions
        & 106×105×21 mm
        & 400×350×70 mm \\

        Weight
        & 0.26 kg
        & 7.3 kg \\

        Price (est.)
        & $\approx$\euro{}3.5k
        & \textgreater\euro{}50k \\
        
        Piezo elements
        & 64
        & 128 \\

        Frequency range
        & 6–15 MHz
        & 7–15 MHz \\

        Max depth
        & 45 mm
        & 30 mm \\

        Scan width
        & 25 mm
        & 23 mm \\

        Raw image access
        & Yes
        & No \\

        RF signal access
        & No
        & No \\
        \bottomrule
    \end{tabular}
\end{table}

\subsubsection{Acquisition setup}
\begin{figure}[tbp]
\centering
\includegraphics[width=0.99\textwidth]{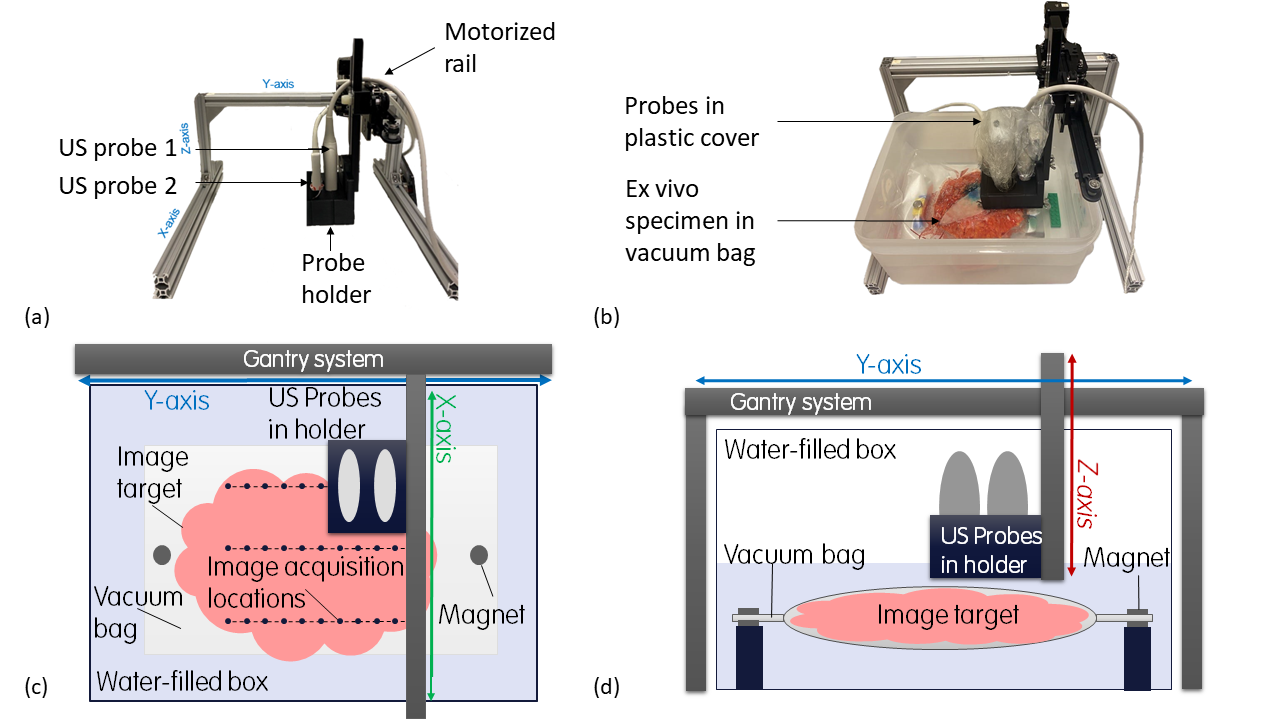}
\caption{Overview of the setup used for paired data collection. The two probes are inserted in the probe holder. The z-axis can be adjusted for the specimen height manually and the x- and y-axis are controlled using the step motor.} 
\label{fig:setup_image}
\end{figure}

A custom motorized acquisition setup was developed to ensure consistent imaging positions across both probes (Fig.~\ref{fig:setup_image}a). A modified two-axis gantry system allowed precise probe translation across predefined grid points, while a 3D-printed holder secured the probes in fixed positions. This setup enabled automated acquisition of paired images at matching locations for both devices, minimizing positional variation.

\subsubsection{Image alignment and registration}
\begin{figure}[tbp]
\centering
\includegraphics[width=1\textwidth]{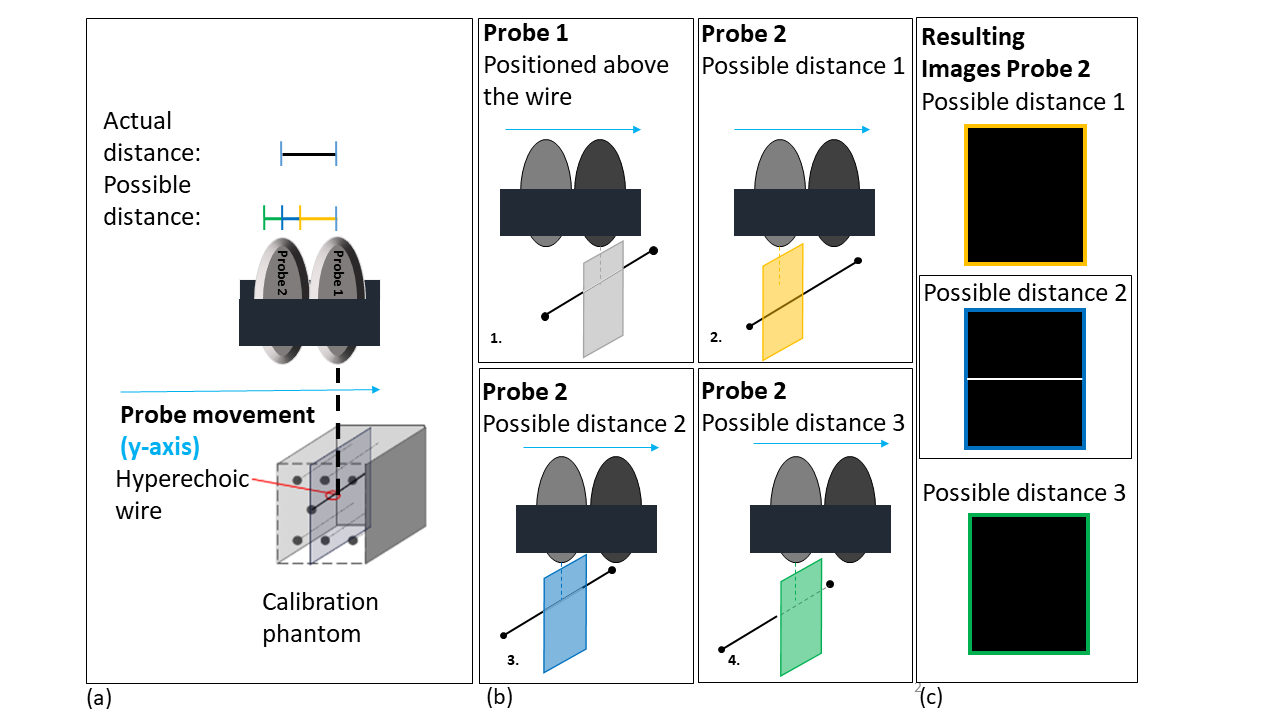}
\caption{Overview of the phantom-based image alignment procedure in the y-direction, performed prior to each measurement session. 
a) Both probes are mounted in the custom probe holder. 
b) The first probe is positioned above the hyperechoic wire target, after which multiple candidate offsets in the y-direction are evaluated for the second probe. 
c) The offset yielding the best visual correspondence of the hyperechoic wire is selected as the optimal distance for the measurement.} 
\label{fig:alignment}
\end{figure}

To further minimize any residual alignment differences between probes, a calibration phantom with hyperechoic wire targets was used. Initial offsets in the y-direction, introduced by slightly different probe positioning in the holder, were corrected by aligning the first probe with the hyperechoic wire target. Subsequently, multiple candidate offsets were evaluated for the second probe, and the offset yielding the best visual overlap was selected for each measurement session, as illustrated in Fig.~\ref{fig:alignment}. 

\begin{figure}[tbp]
\centering
\includegraphics[width=0.99\textwidth]{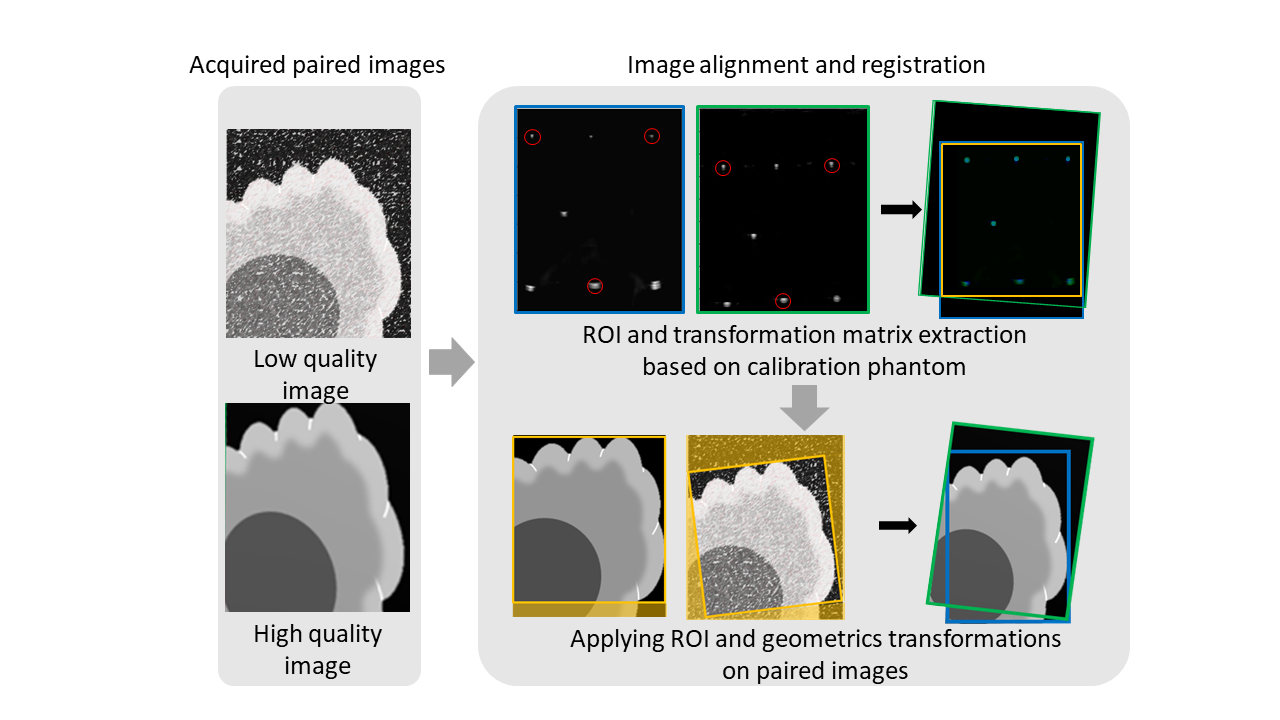}
\caption{Overview of the image alignment and registration workflow applied to the paired dataset. Registration is performed using the calibration phantom images acquired with both the low-quality and high-quality probes. Three corresponding target dots (indicated by red circles) are manually selected in both images, once per measurement session, to compute the necessary registration. These transformations are subsequently applied to all paired low- and high-quality images of the corresponding ex vivo specimen.} 
\label{fig:registration}
\end{figure}

After image alignment, image registration was performed using Elastix with optimized affine transformation parameters (Advanced Normalized Correlation metric, 200 iterations, 40 resolutions, and 2000 spatial samples) to refine spatial correspondence, as demonstrated in Fig.~\ref{fig:registration}. This refinement step ensured accurate pairing of POCUS images and high-end ultrasound images for training deep learning models. 

\subsubsection{Data acquisition and specimens}
Ultrasound images were captured from ex vivo human specimens (breast, colorectal, and sarcoma) and an Abdominal Intraoperative and Laparoscopic
Ultrasound Phantom "IOUSFAN" (Kyoto Kagaku Co., Ltd, Kyoto, Japan). Specimens were placed in vacuum-sealed bags and submerged in water to ensure good acoustic coupling and positional stability (Fig.~\ref{fig:setup_image}b-d). For each specimen, between 40 and 140 images were acquired using a fixed sampling grid (20 mm along the x-axis, 5 mm along the y-axis), reducing slice overlap while capturing distinct anatomical cross-sections.

This study was approved by the Institutional Review Board of the Netherlands Cancer Institute (IRBd26-022). No written consent was required according to the Dutch Medical Research Involving Human Subjects Act (WMO). All patients gave permission in the general hospital consent form  for the use of their data and biological materials for scientific research purposes.

\subsubsection{Image preprocessing}
\begin{figure}[tbp]
\centering
\includegraphics[width=0.99\textwidth]{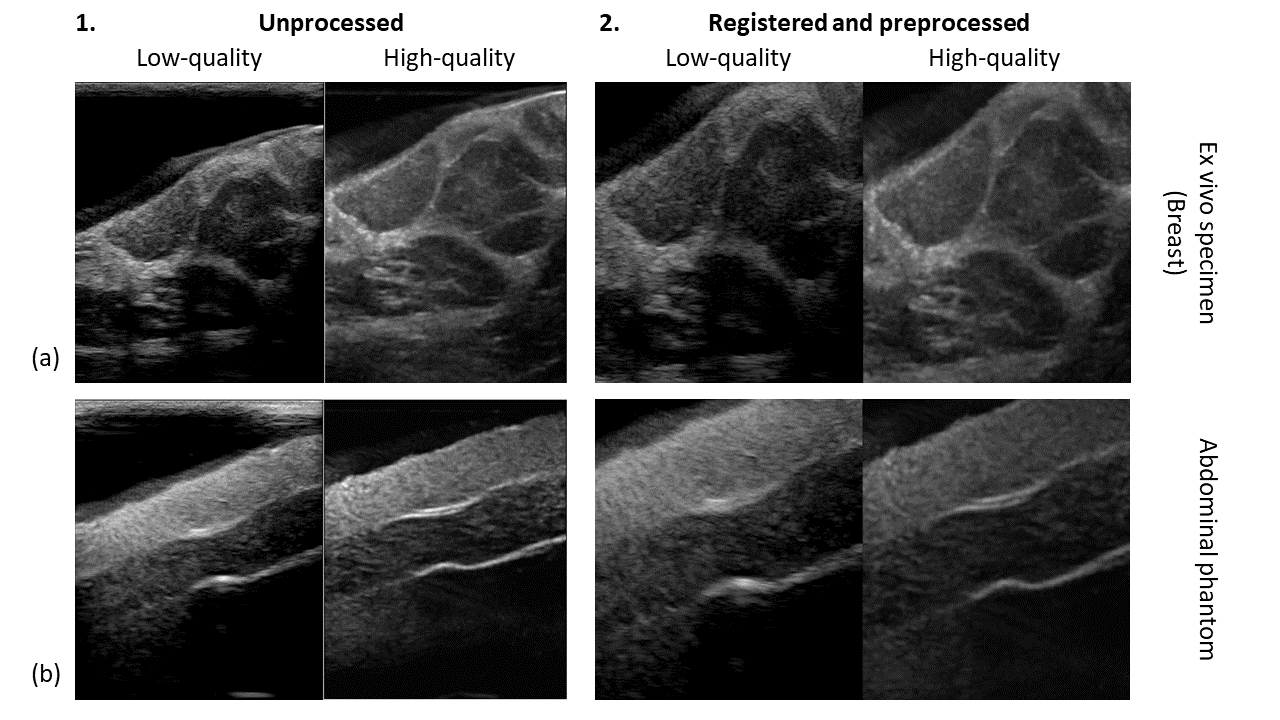}
\caption{Examples of unprocessed and preprocessed ultrasound images acquired with low-cost POCUS (low quality) and a high-end system (high quality) from (a) an excised breast specimen and (b) the abdominal phantom.} 
\label{fig:processed_images}
\end{figure}

Prior to model training, images were preprocessed through cropping, resampling, and normalization. After registration, each low- and high-quality image pair was cropped to their overlapping region to ensure identical physical dimensions. Images were subsequently resampled to consistent sizes, with width and height adjusted to be divisible by 32 while preserving the original aspect ratio. Finally, pixel intensities were normalized to the range [-1, 1] to stabilize network training and improve convergence. Examples of unprocessed and preprocessed images are shown in Fig.~\ref{fig:processed_images}.

\subsubsection{Data availability}
The POCUS-IQ dataset containing paired low-cost POCUS and the high-end cart-based ultrasound images is fully anonymized and publicly available on \url{https://github.com/NKI-MedTech-AI/POCUS-IQ} and Figshare (\url{https://doi.org/10.6084/m9.figshare.31889833}). 

\subsection{Image quality enhancement}
\subsubsection{Deep learning network}
\begin{figure}[htbp]
\centering
\includegraphics[width=0.99\textwidth]{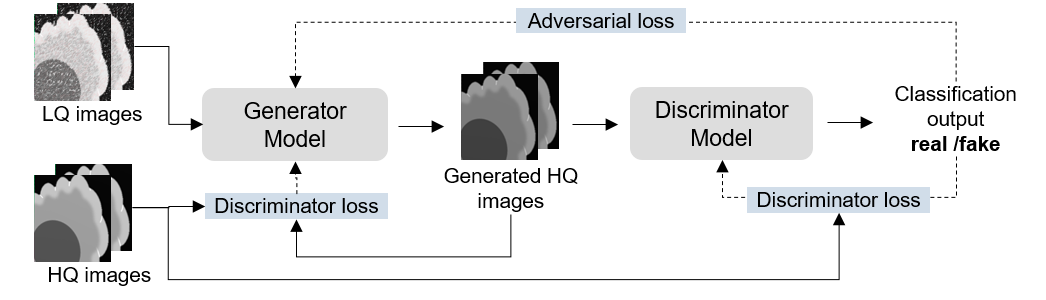}
\caption{Schematic overview of the cGAN framework, consisting of a U-Net generator and a PatchGAN discriminator.} 
\label{fig:model}
\end{figure}

To enhance POCUS image quality, we developed and trained a conditional GAN (cGAN) using our paired dataset of low- and high-quality ultrasound images described above. The network is based on the open-source pix2pix framework proposed by Isola et al. \cite{Isola_16} and consists of a U-Net \cite{ronneberger_15} generator and a PATCHGan discriminator (Fig. \ref{fig:model}).\cite{Isola_16} The U-Net architecture captures both local and global image features by combining downsampling and upsampling paths with skip connections. The PATCHGan discriminator evaluates image realism at the patch level, which promotes the recovery of high-frequency details \cite{Isola_16,gatys_15,gatys_16}.

\subsubsection{Loss function}
Training of the cGAN employed a composite loss function consisting of an adversarial loss~\cite{goodfellow_14}, an L1 pixel-wise loss, and a Structural Similarity Index (SSIM) loss.~\cite{Zhou_19} The adversarial loss encourages realistic image generation, while the L1 and SSIM losses ensure content and structural similarity. The final generator objective combines these losses with weights \(\lambda_1 = 70\) and \(\lambda_2 = 30\):
\begin{equation}
    G^* = \arg \min_G \max_D \mathcal{L}_{\text{cGAN}}(G,D) + \lambda_1 \mathcal{L}_{L1}(G) + \lambda_2 \mathcal{L}_{\text{SSIM}}(G).
\end{equation}

\subsubsection{Network training}
To initialize the network with domain-relevant features, pretraining was performed using a synthetic dataset created by degrading high-quality ultrasound images from multiple open-source datasets \cite{lin_22,al_20,ardakani_23,gomez_24,huang_23,vallez_24,pawlowska_24}. Applied degradations included speckle noise, blurring, compression artifacts, and geometric distortions, designed to mimic common POCUS-related image quality challenges \cite{Zhou_19,salimi_22,Khan_21}. 

The complete dataset, including phantom and ex vivo samples, was split into 90\% training and 10\% test sets. A fixed set of 10 validation images was used to monitor training convergence. All models were implemented in PyTorch (v2.3.1) and trained on an NVIDIA GeForce GTX 1080 GPU using the Adam optimizer (learning rate 0.0002, \(\beta_1=0.5\), \(\beta_2=0.999\)) for 300 epochs. A smaller generator architecture (15k parameters) was used to prevent overfitting and reduce computation time. Data augmentation was applied during training, introducing variation in brightness and contrast, as well as geometric distortions. 

Training progress was monitored through SSIM and L1 losses, quantitative performance metrics, and visual inspection. Due to the adversarial nature of GAN training~\cite{Goodfellow_16}, visual evaluation remained essential to verify that generated high-quality features were structurally consistent with the corresponding low-quality input images. The generator could potentially introduce high-quality features to deceive the discriminator into classifying the image correctly, despite those features not being present in the low-quality image.
An ablation study was performed to assess the impact of adding SSIM loss, pretraining, and data augmentation. 

\subsubsection{Quantitative performance evaluation}
Model performance was assessed using both full-reference and no-reference image quality metrics.

\begin{itemize}
\item Structural Similarity Index Measure (SSIM):  
a full-reference metric that evaluates perceptual similarity based on luminance, contrast, and structure, with constants \(C_1\) and \(C_2\) for numerical stability \cite{Wang_04,hore_10,Sara_19}:
\[
\text{SSIM}(f, g) = \frac{(2\mu_f \mu_g + C_1)(2\sigma_{fg} + C_2)}{(\mu_f^2 + \mu_g^2 + C_1)(\sigma_f^2 + \sigma_g^2 + C_2)},
\]

\item{Peak Signal-to-Noise Ratio (PSNR)}:  
a full reference-metric that measures fidelity using the mean squared error (MSE):
\[
\text{PSNR}(f, g) = 10 \cdot \log_{10} \left( \frac{\text{MAX}_I^2}{\text{MSE}(f, g)} \right).
\]

\item{Naturalness Image Quality Evaluator (NIQE)}:  
a no-reference metric that evaluates perceptual image quality using natural scene statistics. Lower values indicate better quality \cite{Mittal_13}.

\item{Perception-based Image Quality Evaluator (PIQE)}:  
a no-reference perceptual quality metric returning a score between 0 (excellent) and 100 (poor) \cite{Venkatanath_2015}. 
\end{itemize}

\subsubsection{Qualitative performance evaluation}
Besides the quantitative evaluation, a qualitative evaluation was performed to account for perceptual improvements that may not be fully captured by numerical metrics. Images generated by the best-performing model were compared to the corresponding low-quality input images by six researchers with experience in ultrasound image acquisitions and analysis. The model-generated images and the low-quality images were presented side-by-side in random order, while the corresponding high-quality image was simultaneously shown as a ground truth reference. Observers were asked to independently select the image that most closely resembled the high-quality reference. The percentage of selections favoring the model-generated image was reported across all observers and image pairs.

\section{Results}
\label{sect:results}
\subsection{Dataset}
Ultrasound images were acquired using paired low- and high-quality probes across 15 surgical specimens (breast, sarcoma, colorectal) and 1 abdominal phantom, yielding 1299 sets of paired images in total. After excluding image pairs with minimal target content or severe artifacts, 1064 paired image sets remained: 936 from ex vivo surgical specimens and 128 from the phantom. An overview of the final dataset composition is shown in Table~\ref{tab:image_targets}.

\begin{table}[tbp]
    \centering
    \small
    \caption{Overview of the paired ultrasound image dataset, after exclusion of low-content and artefact-contaminated images.}
    \label{tab:image_targets}
    \begin{tabular}{@{}p{2.5cm} p{3.5cm} c c@{}}
        \toprule
        \textbf{Target type} & \textbf{Specifications} & \textbf{Number of targets} & \textbf{Number of image sets} \\
        \midrule
        Ex vivo & Breast specimens & 9 & 527 \\
                & Sarcoma specimens & 3 & 153 \\
                & Colorectal specimens & 3 & 256 \\
        Phantom & IOUSFAN & 1 & 128 \\
        \midrule % Additional line before the total row
        \textbf{Total} & \textbf{Ex vivo \& phantom} & \textbf{16} & \textbf{1064} \\
        \bottomrule
    \end{tabular}
\end{table}

\subsection{Quantitative assessment}
Four model variants were evaluated in the ablation study: a baseline model trained with only L1 loss, a model trained with combined L1 and SSIM loss, a model with the combined loss and additional data augmentation, and a model with the combined loss which was initialized through synthetic pretraining followed by L1+SSIM training. Training each model required approximately 4.5 hours on an NVIDIA GTX 1080 Ti GPU, while pretraining required an additional 25 hours. Inference time on an Intel Xeon W-1250P CPU, limited to 4 cores and threads to emulate deployment on a low-end CPU, was 16.1 ± 1.5 milliseconds per image. All quantitative results are summarized in Table~\ref{own_study_metrics}.

\begin{figure}[tbp]
    \centering
    \includegraphics[width=0.95\textwidth]{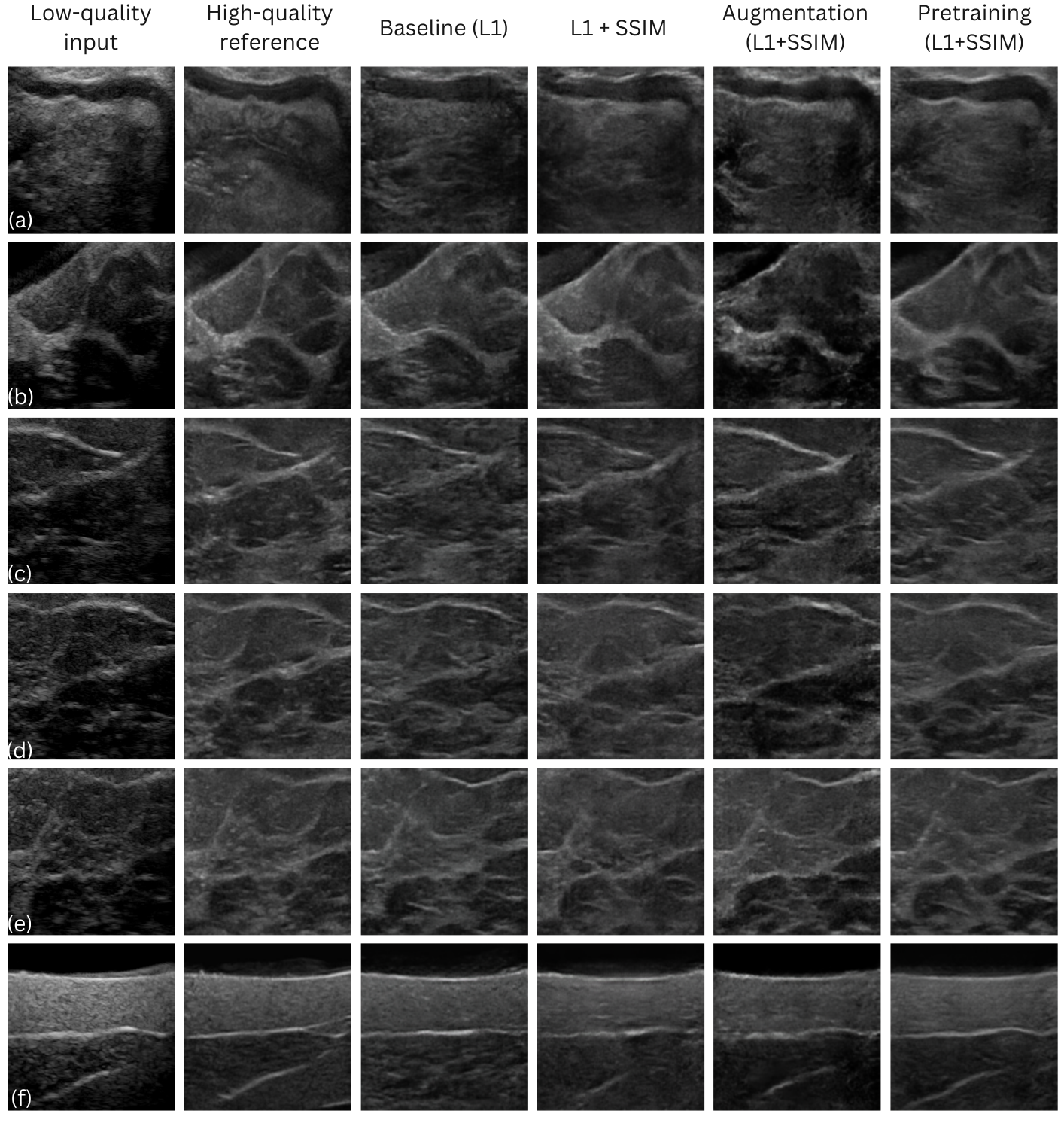}
    \caption{Example images from the test set (a-f), displaying the low-quality input, high-quality reference, and generated output images by each of the four models: Baseline (L1), L1 + SSIM, L1 + SSIM + Augmentation, and L1 + SSIM + Pretraining.}
    \label{generated_images}
    \vspace{2em}
\end{figure}

The Pretraining model achieved the best overall performance, with statistically significant improvements using the paired t-test across all image quality metrics (p \textless 0.001) compared to both the low-quality input and the baseline model. Specifically, SSIM increased from 0.29 to 0.54, PSNR from 19.15 dB to 22.41 dB, while NIQE and PIQE decreased from 7.95 to 4.44 and 31.12 to 19.99, respectively. Visual inspection confirmed that images generated by the Pretraining model more closely resembled the high-quality reference images than those produced by the baseline and augmentation models. This can be seen in (Fig.~\ref{generated_images}), where in row (f) the high-quality reference shows a hyperechoic line in the lower part of the image, which fades away in all but the Pretraining model. Moreover, the upper compartments in row (b) are best preserved in the Pretraining model compared to all the others when compared to the high-quality reference image. 

Consistent performance gains were observed across images acquired from different specimen types, including phantom data and multiple ex vivo tissue types. This suggests that the model effectively learned domain-agnostic features for ultrasound image enhancement.

\begin{table}[tbp]
    \small
    \centering    
    \caption{Quantitative image quality enhancement performance for all evaluated models in this study.}
    \label{own_study_metrics}
    %\resizebox{\textwidth}{!}{%
    \begin{tabular}{lccccc}
        \hline
         \textbf{Method}&\textbf{SSIM} & \textbf{PSNR (dB)} & \textbf{NIQE} & \textbf{PIQE} \\
        & (mean $\pm$ std) & (mean $\pm$ std) & (mean $\pm$ std) & (mean $\pm$ std) \\
        \hline
        {High-quality reference} & - & - & $5.17 \pm 0.60$ & $23.29 \pm 6.74$ \\
        {Low-quality input} & $0.29 \pm 0.06$ & $19.15 \pm 1.95$ & $7.95 \pm 1.78$ & $31.11 \pm 5.91$ \\
        {Baseline (L1)} & $0.49 \pm 0.09$ & $21.74 \pm 2.00$ & $5.13 \pm 0.46$ & $28.98 \pm 4.53$ \\
        {L1 + SSIM} & $0.52 \pm 0.07$ & $22.07 \pm 2.09$ & $5.33 \pm 0.71$ & $28.26 \pm 5.83$ \\
        {L1 + SSIM + Augmentation} & $0.44 \pm 0.09$ & $20.41 \pm 2.28$ & $5.56 \pm 1.24$ & $34.99 \pm 7.64$ \\
        {L1 + SSIM + Pretraining} & $0.54 \pm 0.08$ & $22.41 \pm 2.19$ & $4.44 \pm 0.53$ & $19.99 \pm 5.72$ \\
        \hline
    \end{tabular}
    %}
\end{table}

\subsection{Qualitative assessment}
In addition to quantitative evaluation, a qualitative assessment was performed to evaluate perceptual image quality improvements. A randomly selected subset of the test dataset was independently reviewed by six researchers with a  technical-medicine or biomedical engineering background, which were asked to select the image that most closely resembled the high-quality reference. The results are summarized in Table \ref{qualitative_assessment}. Across all evaluations and reviewers, the model-generated images were preferred in 75\% of comparisons, while in the remaining 25\% the original low-quality images were favored. These results indicate that the proposed image enhancement method improved perceived image quality in the majority of cases, better resembling the ground truth.

%\begin{table}[htbp]
%    \centering
%    \small
%    \caption{Win percentages from the qualitative assessment with six observers, %comparing low-quality input images and model-generated images.}
%    \label{qualitative_assessment}
%    \begin{tabular}{lc}
%        \toprule
%        & \textbf{Win percentage (\%)} \\
 %       \midrule
 %       {Low-quality images} & 25 \\
 %       {Model-generated images} & 75 \\
 %       \bottomrule
 %   \end{tabular}
%\end{table}

\begin{table}[htbp]
    \centering
    \small
    \caption{Win percentages from the qualitative assessment with six observers, comparing low-quality input images and model-generated images.}
    \label{qualitative_assessment}
    \resizebox{\linewidth}{!}{
    \begin{tabular}{lccccccc}
        \toprule
        & Reviewer 1 & Reviewer 2 & Reviewer 3 & Reviewer 4 & Reviewer 5 & Reviewer 6 &Overall percentage (std)\\
        \midrule
        {Low-quality image} & 17\% & 17\% & 50\% & 42\% & 17\% & 8\% & 25\% $\pm$ 14\% \\
        {Model-generated images} & 83\% & 83\% & 50\% & 58\% & 83\% & 92\% & 75\% $\pm$ 15\% \\
        \bottomrule
    \end{tabular}
    }
\end{table}

\section{Discussion}
\label{s:discussion}
\subsection{Key findings and contributions}
This study introduces a novel accurately paired ultrasound image dataset (POCUS-IQ) acquired using a low-cost POCUS device and a high-end ultrasound system, which has been made publicly available to support benchmarking and future methodological development. A total of 1064 paired ex vivo and phantom image sets were obtained using a motorized acquisition setup that ensured consistent probe positioning across devices, followed by a dedicated alignment and registration pipeline. This unique dataset allowed for the development and validation of an image-to-image enhancement model based on a conditional GAN (cGAN) with a U-Net generator, pre-trained on synthetically degraded ultrasound data.

The proposed model achieved substantial improvements in image quality across all evaluated metrics, increasing SSIM and PSNR by 86\% and 17\%, and reducing NIQE and PIQE by 44\% and 36\%, respectively. Notably, the NIQE and PIQE scores of the enhanced images surpassed those of the high-end reference device. Qualitative evaluation further confirmed that the generated images more closely resembled the high-quality reference images than the original low-quality POCUS inputs. The model’s short inference time on CPU (16.1 ms per image) underscores its potential for real-time applications on low-power mobile devices without a powerful GPU.

%When compared to prior studies such as Zhou et al. \cite{Zhou_19,Zhou_21}, Guo et al.\cite{Guo_20}, and the USenhance dataset~\cite{miccai_23}, our approach demonstrated superior absolute performance, despite a relatively smaller number of unique specimens (Table~\ref{combined_metrics}). This discrepancy can be partially attributed to the use of accurately registered paired data in our study, which is often lacking in other works. Many earlier studies report substantial improvements starting from very low baseline quality metrics, potentially inflating perceived performance gains. In contrast, our study offers a more reliable benchmark for enhancement performance by using a well-aligned dataset.

\subsection{Comparison with literature}
A recent literature review by the authors highlighted substantial variability in reported low-cost input quality metrics and performance gains for POCUS image enhancement across studies~\cite{Pol_24}. These differences are likely due to heterogeneity in dataset composition, ultrasound devices, and alignment quality. Additionally, an important observation from this review was that simulation-based datasets tend to yield larger performance improvements than in vivo or phantom datasets, suggesting that simulated data may not fully capture the complexity and variability of clinical scenarios. Therefore, our results can best be compared to studies with clinical in vivo or ex vivo datasets.

%An additional strength of our approach lies in the use of both full-reference and no-reference image quality metrics to assess performance, enabling a holistic understanding of model behavior. While the SSIM and PSNR metrics reflect the structural and pixel-wise improvements achieved, the no-reference NIQE and PIQE metrics help evaluate perceptual quality, often aligning better with clinical intuition. The combination of these metrics provides a robust evaluation framework. 

Many earlier studies report large relative improvements that originate from very low baseline image quality metrics, potentially inflating perceived performance gains. For example, baseline PSNR values generally ranged from 8.65 to 16.04 in literature~\cite{Zhou_18,Zhou_19,Zhou_21,Guo_20} compared to 19.16 in this study, while the baseline SSIM ranged from 0.18 to 0.24~\cite{Zhou_19,Zhou_21} compared to 0.29 in this study. One study of Moinuddin et al.~\cite{Moinuddin_2022} did report higher baseline quality (PSNR 26.01, SSIM 0.71), resulting in only minimal relative improvements (+3.5\% PSNR, +6.0\% SSIM). In contrast, our proposed method achieves both substantial relative gains and strong absolute performance, supported by a well-aligned paired dataset that enables more reliable benchmarking.

In literature, the reconstructed images have a SSIM and a PSNR range of 0.41 to 0.75 and 18.08 to 26.91 respectively, as can be seen in Table \ref{combined_metrics}. However, it is important to point out that Moinuddin et al. used the BUS dataset, which did not use POCUS to acquire the images.\cite{Yap_2018} The baseline quality of the BUS dataset could be higher than a POCUS device, thus leading to a higher reconstructed SSIM and PSNR. Therefore, the SSIM of 0.54 and PSNR of 22.41 reported in this paper are among the highest absolute performance metrics. Moreover, while the absolute metrics of Moinuddin et al. are comparatively high, the relative increase for both the SSIM and the PSNR are low, reaching an improvement of +6.0\% and +3.5\% respectively.\cite{Moinuddin_2022} Table \ref{combined_metrics} does not contain any metrics regarding the NIQE or PIQE, since these metrics were not reported in any of the studies. 

Besides the results from Moinuddin et al., the results in this research have an absolute higher value than those of the other studies mentioned in Table \ref{combined_metrics}. Furthermore, studies with a low baseline SSIM and PSNR reported large relative improvements.\cite{Zhou_18,Zhou_19,Zhou_21,Guo_20} Because these results originated from a low baseline, the performance gain could potentially be inflated. The reason for our increased baseline performance metrics might be due to the difference in the dataset. In contrast to the aforementioned studies, our study benefits from an optimally aligned dataset, where the baseline images have minimal spatial misalignments. Therefore, the baseline images might resemble the high-quality reference images better. 

\begin{table}[tbp]
    \small
    \centering    
    \caption{Quantitative image quality enhancement performance for all evaluated models in this study and comparison with previously reported methods in literature.}
    \label{combined_metrics}
    %\resizebox{\textwidth}{!}{%
    \begin{tabular}{llccc}
        \hline
         \textbf{Study}&\textbf{Data}&\textbf{SSIM} & \textbf{PSNR (dB)} &\\
        && (mean $\pm$ std) & (mean $\pm$ std) & \\
        \hline
        This study&{High-quality reference} & - & - & \\
        &{Low-quality input} & $0.29 \pm 0.06$ & $19.15 \pm 1.95$ &\\
        &{L1 + SSIM + Pretraining} & $0.54 \pm 0.08$ & $22.41 \pm 2.19$ &  \\
        \hline

         Guo et al. \cite{Guo_20} 
        & Low-quality input & - & 16.04 &\\
        & Reconstructed images & - & 18.94& \\
        Zhou et al. \cite{Zhou_19} 
        & Low-quality input & 0.18 $\pm$ 0.04 & 8.65 $\pm$ 1.32 &\\
        & Reconstructed images & 0.41 $\pm$ 0.05 & 18.08 $\pm$ 1.57&\\
        Zhou et al. \cite{Zhou_21} 
        & Low-quality input & 0.24 $\pm$ 0.06 & 12.68 $\pm$ 3.45& \\
        & Reconstructed images & 0.45 $\pm$ 0.06 & 19.95 $\pm$ 3.24& \\
        Muonduddin et al.\cite{Moinuddin_2022}
        & Low-quality input & 0.71 $\pm$ 0.08 & 26.01 $\pm$ 2.31 &\\
        & Reconstructed images & 0.75 $\pm$ 0.06 & 26.91 $\pm$ 2.30 &\\
        \hline
        
    \end{tabular}
    %}
\end{table}

An additional strength of our approach lies in the use of both full-reference and no-reference image quality metrics to assess performance. While the SSIM and PSNR reflect the structural and pixel-wise improvements achieved with respect to the reference images, the no-reference NIQE and PIQE help evaluate perceptual image quality, often more aligned with clinical interpretation. The agreement between these metrics and the qualitative evaluation strengthens confidence in the observed improvements.

%However, several limitations exist. The diversity of ex vivo specimens in the dataset was limited, which may restrict generalizability. The automated scanning occasionally captured images lacking clear anatomical structures, and the test set was relatively small due to the need for a large training set. While image normalization was limited to rescaling intensities, experiments with gray-level stretching and SNV normalization were inconclusive or infeasible. The training process also encountered typical GAN instability, and no cross-validation was implemented due to computational constraints. Although our method outperforms previous research, a PSNR of 22 is still generally considered as moderate. 

\subsection{Limitations}
Despite these promising results, several limitations should be acknowledged. First, the dataset comprises a limited number of ex vivo tissue types (breast, colorectal, and sarcoma), and all acquisitions were performed on ex vivo specimens or phantoms. Moreover, specimens were vacuum-sealed and submerged in water to ensure acoustic coupling, which differs from typical in vivo imaging conditions. As a result, the generalizability of the model to other tissue types and in vivo scenarios remains to be established.

Second, the automated grid-based acquisition, necessary to construct our accurately paired dataset, occasionally captured images with limited anatomical content and mostly water. As mentioned in the methods section, those image pairs (18\% of the total acquired data) were excluded since images without clear structures could have negatively impacted model training. To compensate, a large proportion of the dataset was allocated to training, resulting in a relatively small test set. 

From a methodological perspective, image normalization was restricted to intensity rescaling, as more advanced techniques such as gray-level stretching or standard normal variate normalization yielded inconclusive results. Training instability inherent to GAN-based approaches was also observed, consistent with prior reports~\cite{mescheder_18,ahmad_24}. Moreover, no cross-validation was performed due to computational constraints, which, combined with small performance differences in the ablation study, limited definitive selection of an optimal configuration. While the proposed model outperforms prior work, a PSNR of approximately 22~dB is still considered moderate, indicating room for further improvement.
%Future work should explore improved normalization techniques, more diverse specimen inclusion, and increased test data. The model architecture could be enhanced with perceptual loss functions such as VGG-based metrics or domain-specific variants, and more advanced architectures like residual or two-stage GANs may offer benefits, albeit with increased computational costs. Moreover, the dataset enables validation of enhanced POCUS for real-time applications, such as diagnostic imaging in general practice, emergency rooms, or image-guided interventions.

\subsection{Future directions}
Future work should focus on expanding the dataset in both size and diversity, including additional tissue types and, critically, in vivo acquisitions. This would enable the creation of a larger test set as well, facilitating more robust and reliable model validation. Moreover, enhanced POCUS imaging could be used for a large amount of real-time clinical applications, including diagnostic imaging in general practice, emergency care, and image-guided interventions. Consequently, it remains essential to investigate how the model performs in in vivo clinical settings representing the intended end-use case.

Further improvements in model performance may be achieved through advances in both normalization strategies and network architecture. More sophisticated normalization techniques could help direct the model’s focus toward anatomically relevant structures rather than speckle noise. In addition, the architecture could be enhanced by incorporating perceptual loss functions, such as VGG-based or domain-specific feature metrics. More advanced designs, including residual or two-stage GAN architectures, may further improve image quality, albeit at the cost of increased computational complexity. 

Ultimately, the value of image enhancement lies not only in improved visual quality but in its ability to support clinical decision-making and downstream tasks such as lesion localization, segmentation, or classification. As handheld ultrasound devices continue to become more accessible and affordable, robust post-processing methods such as the one proposed in this study may facilitate broader adoption of POCUS across healthcare systems, particularly in point-of-care environments and resource-limited countries.

\section{Conclusions}
This study presents the first accurately paired dataset of low-quality POCUS and high-end ultrasound images, crucial for advancing post-image enhancement methods for POCUS. This publicly available dataset contains 1064 image pairs acquired from breast, colorectal, and sarcoma tissue, and facilitates model development and reliable validation with paired, accurately registered high-quality reference images. 
Using this dataset, a conditional GAN-based image enhancement model was developed. The model achieved significant improvements in image quality over the low-quality POCUS input images, with performance metrics surpassing those reported in related studies. The model’s real-time inference capability on CPU further highlights its suitability for deployment on portable and low-power systems.
While additional validation in in vivo and clinical settings is required, this work represents an important step toward reliable, data-driven image quality enhancement for POCUS. By preserving the inherent advantages of affordability and portability, the proposed approach has the potential to increase the value and impact of POCUS in both high-resource and resource-limited healthcare environments.

% \disclosures 
\subsection*{Disclosures}
The authors declare no conflicts of interest.

\subsection*{Code and Data Availability}
The POCUS-IQ dataset underlying the results presented in this paper are publicly available and can be located on Figshare (\url{https://doi.org/10.6084/m9.figshare.31889833}).

\subsection* {Acknowledgments}
This work was supported by the Dutch Cancer Society (KWF Smart Measurement Technologies - 15665). We are grateful for their support. Research at the Netherlands Cancer Institute is supported by institutional grants of the Dutch Cancer Society and of the Dutch Ministry of Health, Welfare and Sport.

%%%%% References %%%%%

\bibliography{mybibliography}   % bibliography data in report.bib
\bibliographystyle{spiejour}   % makes bibtex use spiejour.bst

%%%%% Biographies of authors %%%%%

\vspace{2ex}\noindent\textbf{Lennard M. van Karnenbeek} is a PhD candidate at the University of Twente and the Netherlands Cancer Institute in Amsterdam. He received his Masters degree at the University of Twente in Technical Medicine in 2022.

\vspace{1ex}
\noindent Biographies and photographs of the other authors are not available.

\listoffigures
\listoftables

\end{spacing}
\end{document}